\documentclass[dvips]{acta}
\usepackage{supertabular,lscape,epsfig}
\usepackage{amssymb}
\usepackage{amsmath}

\SetPages{0}{0}

\SetVol{66}{2016}

\begin{document}

\begin{Titlepage}
\Title{High proper motion objects towards the inner Milky Way: characterisation of newly identified nearby stars from the VISTA Variables in the V\'ia L\'actea Survey}

\Author{G~r~o~m~a~d~z~k~i$^{1,2}$, M., K~u~r~t~e~v$^{2,1}$, R., B~e~a~m~\'i~n$^{1,2}$, J.~C.,
T~e~k~o~l~a$^{3}$, A., R~a~m~p~h~u~l$^{4,5}$, R., I~v~a~n~o~v$^{6}$, V.~D., M~i~n~n~i~t~i$^{7,1,8}$, D., 
F~o~l~k~e~s$^{9,2}$, S.~L., V~a~i~s~a~n~e~n$^{4,10}$, P., K~n~i~a~z~e~v$^{4,10,11}$, A.~Y., 
B~o~r~i~s~s~o~v~a$^{2,1}$,~J., P~a~r~s~o~n~s$^{12}$, S.~G. and V~i~l~l~a~n~u~e~v~a$^{1,2}$, V.}
{$^{1}$Millennium Institute of Astrophysics, Av. Vicua Mackenna 4860, 782-0436, Macul, Santiago, Chile\\
$^{2}$Instituto de F\'isica y Astronom\'ia, Universidad de Valpara\'iso, Av. Gran Breta\~{n}a 1111, Playa Ancha, Casilla 5030, Valpara\'iso, Chile\\
e-mail:mariusz.gromadzki@uv.cl\\
$^{3}$Las Cumbres Observatory Global Telescope Network, Inc., 6740 Cortona Drive, Suite 102, Goleta, CA  93117, USA\\
$^{4}$South African Astronomical Observatory, P.O. Box 9 7935, South Africa\\          
$^{5}$University of Cape Town, Astronomy Department, Rondebosch 7701, South Africa\\
$^{6}$European Southern Observatory, Karl-Schwarzschild-Str. 2, 85748 Garching bei M\"unchen, Germany\\
$^{7}$Departamento de Ciencias F\'isicas, Universidad Andres Bello, Republica 220, Santiago, Chile\\
$^{8}$Vatican Observatory, V00120 Vatican City State, Italy\\
$^{9}$Centre for Astrophysics Research, Science and Technology Research Institute, University of Hertfordshire, Hatfield AL10 9AB, UK\\
$^{10}$Southern African Large Telescope Foundation, P.O. Box 9 7935, South Africa\\
$^{11}$Sternberg Astronomical Institute, Lomonosov Moscow State University, Moscow, Russia\\
$^{12}$Department of Physics and Astronomy, University of Sheffield, Sheffield S3 7RH, UK}

\Received{June 10, 2016}
\end{Titlepage}

\Abstract{The census of the Solar neighbourhood is still incomplete, as demonstrated by recent discoveries of many objects within 5--10\,pc from the Sun. The area around the mid-plane and bulge of the Milky Way presents the most difficulties in searches for such nearby objects, and is therefore deficient in the known population. This is largely due to high stellar densities encountered.
 Spectroscopic, photometric and kinematic characterization of these objects allows better understand the local mass function, the binary fraction, and provides new interesting targets for more detailed studies.
 We report the spectroscopic follow-up and characterisation of 12 bright high PM objects, identified from the VISTA Variables in V\'ia L\'actea survey (VVV). We used  the 1.9-m telescope of the South African Astronomical Observatory (SAAO) for low-resolution optical spectroscopy and spectral classification, and the MPG/ESP 2.2m telescope Fiber-fed Extended Range Optical Spectrograph (FEROS) high-resolution optical spectroscopy to obtain the radial and space velocities for three of them.
 Six of our objects have co-moving companions. We derived optical spectral types and photometric distances, and classified all of them as K and M dwarfs within 27 -- 264\,pc of the Sun. Finally, we found that one of the sources, VVV\,J141421.23-602326.1 (a co-moving companion of VVV\,J141420.55-602337.1), appears to be a rare massive white dwarf that maybe close to the ZZ\,Ceti instability strip. Many of the objects in our list are interesting targets for exoplanet searches.}
{proper motions -- stars: low-mass -- (stars:) white dwarfs -- (stars:) binaries: visual -- (Galaxy:) solar neighbourhood -- techniques: spectroscopic.
}

\section{Introduction}
M-dwarfs account for over 70\% of stellar systems in the solar vicinity 
(Henry \etal 1997). Most of them are single 
($\sim$60-70\%; Fischer \& Marcy 1992; Bergfors \etal 2010), making them more 
likely to host (potentially habitable)  planets
(e.g. Kraus \etal 2012). For comparison, 
the single star fraction is $\sim$54\% for solar-type stars (Duquennoy \& Marcy 1991; Raghavan \etal 2010) and 
it is $\sim$0\% for massive stars (Preibisch \etal 1999). 
This makes M-type dwarf stars the most numerous potential planet hosts of all 
the stellar classes (Lada 2006). 
Furthermore, all exoplanet detection methods (radial velocity, transits, 
direct imaging with Adaptive Optics and astrometry) are more sensitive to 
planets with host stars of lower masses. A number of exoplanet search programs 
are aggressively targeting M-dwarfs with radial velocities ({\it e.g.} M2K; Apps \etal 2010) and with transits 
(RoPACS and MEarth; Irwin \etal 2014).  The first 
exoplanet to be imaged was orbiting a brown dwarf (BD) at $\sim 70$\,pc 
(Chauvin \etal 2005), and there is a on-going debate for an astrometrically 
discovered third planetary mass body in a nearby BD pair at 
$\sim 2.3$\,pc (Boffin \etal 2014; Sahlmann \& Lazorenko 2015).

One of the most powerful methods to identify nearby stars is through identifying proper motion (PM), and by exploiting 
a PM search at infrared wavelengths, offers an additional advantage: cool objects are intrinsically brighter at those 
wavelengths than in the optical because their spectral energy distributions peak at $\lambda$$>$1\,$\mu$m (L\'epine \& Gaidos 2011) 
published an all-sky catalog of M-dwarfs with apparent near-InfraRed (near-IR) magnitude $J$$<$10. They selected 8889 stars 
from the on-going SUPERBLINK ({\it e.g.} L\'epine \& Shara 2005) survey of stars with $\mu$$>$40\,mas\,yr$^{-1}$, supplemented 
at the bright end with the TYCHO-2 catalogue. Recently, Lepine \etal (2013) presented a spectroscopic catalog of the 1564 
brightest ($J$$<$9) M-dwarf candidates in the northern sky.

The majority of surveys avoid Galactic plane and bulge, or are substantially incomplete near to these regions. 
However, these regions offer considerable latent potential for new discoveries of nearby low-mass stars and brown dwarfs. 
This is especially true for nearby or bright examples that have been overlooked in previous searchs due to confusion 
caused by high stellar densities and background contaminant objects (Folkes \etal 2012; Luhman 2013; Scholz 2014). 
A fortuitous aspect of discoveries at low Galactic latitudes  is that these regions typically offer many suitable 
reference stars for high-Strehl ratio AO follow-up observaitions.

This is the third paper of a project, after Beam\'in \etal (2013) and Ivanov \etal (2013), to generate a uniform catalog of
high-PM objects within the VVV footprint, and characterise them 
with spectroscopic follow-up observations.
It is organised as follows: the next section describes the sample selection, and 
the new observations. 
Section 3 describes the spectral type estimation, the distance 
measurements, and reports on the co-moving companions. Finally, 
in section 4 we present summary and conclusions.

\section{Sample selection and observations}

The targets reported here were identified during a test phase of 
our method to search for high PM objects using the VVV survey 
database. The VVV observations at the time of the search covered 
the period from Feb 2010 to Mar 2013.  
Our PM analysis techniques makes use of four $K_{\rm S}$-band 
epochs, preferentially selected with equally spaced epochs, obtained under similar 
seeing and photometric conditions.
We used the source catalogs generated by the Cambridge 
Astronomical Survey Unit 
(CASU)\footnote{http://casu.ast.cam.ac.uk/surveys-projects/vista}.
Their astrometric precision is 0.05--0.09\,arcsec. Pairs of catalogs 
for neighbouring epochs were cross-identified with a matching radii 
scaled to correspond to a maximum PM of 5\,arcsec\,yr$^{-1}$ using 
the STILTS code (Taylor 2006). 
This procedure was repeated for all sequential pairs of epochs, yielding
three PM measurements for each object.
Next, we removed candidates with inconsistent PMs. 
At this stage of the project we were interested in nearby
objects with larger apparent motions, so we imposed two additional constraints selecting only 
brightest stars with $K_{\rm S}$$<$13.5\,mag, and with PMs exceeding 3 times 
the astrometric errors. All final candidates were visually inspected.
The proper motions of targets were estimated using the 2MASS and last available VVV epoch
data, that gave a typical error of proper motion equal to 0.01\,arcsec\,yr$^{-1}$.
The resulting catalog and a more detailed description of the search 
will be reported in Kurtev et al. (2016).

For this paper we selected 12 high-PM stars based on their magnitudes 
and the visibility at the moment of the observations. 
Four of our targets have been previously identified as high PM stars: 
VVV~J121051.57-642528.5 and VVV~J164622.06-420118.8 were listed in 
TYCHO-2 catalog (H{\o}g \etal 2000) as TYC 8982-1530-1 and TYC 7875-141-1, 
respectively. VVV~J122701.70-634203.7 and~VVV J141420.55-602337.1 were
reported by (Finch \etal 2010) as 2MASS~J122701.32-634203.1 and 2MASS~J141420.90-602336.1, respectively.

Throughout this paper we adopt a naming convention with the survey 
abbreviation followed by the J2000 coordinates: VVV~JHHMMSS.ss-DDMMSS.s, 
according to the IAU convention.

\subsection{Radcliffe/SAAO spectra}

Low-resolution spectra of 12 stars were acquired on
 2013 Apr 4--7 using the Grating Spectrograph with the SITe
(Scientific Imaging Technologies, INC.) CCD
mounted on the 1.9-m Radcliffe telescope at the South African
 Astronomical Observatory (SAAO), Sutherland. We made use of grating
 no. 7 with 300 lines mm$^{-1}$ and a slit with a projected width of
 1.35 arcsec.
During observations seeing varied from 0.8 to 2.5 arcsec, with an average around 1.5 arcsec. The total duration of an exposure was
 600-1200\,s, depending on object's brightness, split into two individual integrations. 
 Usually, two or three standards were observed on each night,
selected among the list: LTT\,3864, LTT\,4816, LTT\,7379, 
LTT\,6248, and CD$-$32$^{\circ}$~9927.
This allowed us to flux-calibrate the low-resolution spectra in absolute units.
The wavelength calibration was performed
using CuAr reference lamp spectra. All the data reductions and
calibrations were carried out with standard {\sc IRAF}\footnote{{\sc IRAF} is distributed by the National Optical Astronomy Observatory, which is operated by the Association of Universities for Research in Astronomy (AURA) under cooperative agreement with the National Science Foundation.} procedures.
The final spectra from the first two nights cover the range 4292--8289\AA~
and from the last two nights 3761--7735\AA~ with a resolving power of R$\sim$1000.

\subsection{FEROS/ESO spectra}

We obtained radial velocities (RV) for VVV~J121436.36-640808.4,  VVV~J132355.14-620324.9 and VVV~J164810.92-414014.9
with the Fiber-fed Extended Range Optical Spectrograph (FEROS; Kaufer \etal 1999), at the MPG/ESP 2.2m telescope.
The instrument has a resolving power of 48\,000 over the spectral range 350\,-\,920 nm. FEROS uses two fibers of 2 arcsec ,
in diameter with a separation of 2.9 arcmin. One of the fibers was placed on the star and the other was fed by a Thorium-Argon lamp 
to obtain a better wavelength calibration (Object-calibration mode). The observations were carried out on the nights  15 and 16 Feb 2015. 
The seeing conditions for each night were varied between 0.5--1.1 arcsec  and 0.6--1.4 arcsec, respectively.

In order to reduce the data we obtained the standard calibration frame for this instrument (i.e. bias, flat, lamps).
Here we give a brief summary of the reduction process, but the reader is referred to Jord\'an \etal (2014) and 
Brahm, Jord{\'a}n and Espinoza (2016) for a detailed explanation of each step.
First, the bias was removed, and the flat fielding correction was applied. Next, one-dimensional spectra were optimally extracted 
for each echelle order of both the science and the calibration spectra. The wavelength calibration is first processed order by order 
using a reference Thorium-Argon lamp, each order is fitted against this reference  until the R.M.S. is less than 70\,m\,s$^{-1}$ 
then a global solution is performed iteratively until an R.M.S. below 100\,m\,s$^{-1}$ is reached. Then, the solution is applied to the target
spectrum. Finally, a barycentric correction is applied using the Jet Propulsion Lab ephemerides (JPLEphem) package\footnote{https://pypi.python.org/pypi/jplephem}.

The reduced spectra are cross correlated with a set of synthetic spectra  from Coelho \etal (2005) to estimate the
physical parameters in a iterative way, and after convergence a binary mask is used to measure the RV via the cross correlation
function (Baranne \etal 1996).
A simple Gaussian is fitted to the cross correlation function with the binary mask, and the mean is taken as the RV of the star.
The binary masks are the same ones used by the HARPS data reduction (Mayor \etal 2003).

\subsection{VISTA Variables in V{\'i}a L{\'a}ctea survey}

VISTA Variables in the V{\'i}a L{\'a}ctea (VVV) is a public ESO (European Southern Observatory) near-infrared (near-IR) survey 
that is mapping the Milky Way Bulge and an adjacent section of the mid-plane with 
the VISTA telescope (Minniti \etal 2010).

The VISTA is a 4.1-m telescope, located on Cerro Paranal, 
equipped with VIRCAM (VISTA InfraRed CAMera; Dalton \etal 2006), a
wide-field camera producing $\sim$1$\times$1.5 deg$^{2}$ tiles, working in the
0.9--2.4 $\mu$m wavelength range. The VISTA data are processed
with the VISTA Data Flow System (VDFS; Irwin \etal 2004;
Emerson \etal 2004) pipeline at the Cambridge Astronomical
Survey Unit. The data products are available through the ESO
archive or the specialised VISTA Science Archive (VSA; Cross \etal 2012).

The VVV survey covers a total area of 562\,deg$^2$ in the Galactic bulge and southern disk. 
The VVV database now contains multicolour photometry in $ZYJHK_{\rm S}$-bands, and multiple epochs in the $K_{\rm S}$-band, 
monitoring a billion sources in total (Saito \etal 2012). The time baseline already exceeds 5 years, from the first observations acquired 
in Oct 2009, and the plan is to extend this for a
couple more years until the survey is be completed. 
The VVV survey gives unique information on these inner regions of the Milky Way that 
have remained mostly uncharted due to crowding and heavy extinction, 
allowing a number of studies of stellar populations and Galactic structure 
({\it e.g.} Hempel \etal 2014). In particular, the time baseline, combined with the astrometric accuracy of $\sim$25 mas 
for a $K_{\rm s}$$=$15.0 mag 
source and $\sim$175 mas for $K_{\rm s}$$=$18.0 mag enable a number of interesting astrometric studies. 
The typical proper motion measurements for non saturated and approximately bright sources for the first 
four years reach an accuracy of $\sim$10 mas yr$^{-1}$ ($K_{\rm s}$$=$15.0 mag) and $\sim$20 mas yr$^{-1}$ 
($K_{\rm s}$$=$18.0 mag). More details of astrometry with the VVV are given in Saito \etal (2012).
We have just started to exploit the VVV database for proper motion studies: 
Beam\'in \etal (2013) reported the first VVV brown dwarf discovery,  
Ivanov \etal (2013)  found seven new companions to known high PM nearby stars, 
Libralato \etal (2015) produced a high-precision astrometry reduction pipeline for the VVV survey 
data, and  Kurtev \etal (2016) produced a catalogue of 3\,003 high proper motion stars in the VVV fields. 

\subsection{Other photometric data}

We complemented the VVV observations with archival
multi-wavelength photometry to obtain their spectral energy distributions. Photometric data for our objects were found in the TYCHO-2 catalog H{\o}g \etal (2000), 
the Fourth U.S. Naval Observatory CCD Astrograph Catalogue (UCAC4; Zacharias \etal 2012, 2013),
the Deep Near-Infrared Survey of the Southern Sky (DENIS; Epchtein \etal 1997), 
the Two Micron All Sky Survey  (2MASS; Skrutskie \etal 2006), the Wide-Field
Infrared Survey Explorer  (AllWISE; Wright \etal 2010; Cutri \etal 2013) and Spitzer Galactic Legacy Infrared Mid-Plane Surveys Extraordinaire (GLIMPSE; Benjamin \etal 2003). 
The cross-identification of the catalogs was 
carried out with Topcat (Taylor 2006), and we ensured that our high PM 
objects are correctly cross-identified with a visual inspection of the images,
facilitated with the Aladin tool (Bonnarel \etal 2000). The photometric
measurements are available in the online materials.

\section{Results}

\subsection{Spectro-photometric characterisation}

We derived spectral type for our targets using three methods: 
direct comparison with spectral templates, spectral indices, 
and spectral energy distributions (SEDs) fitting. We find good agreement 
between spectral types determined with the first two methods. 
The third method yields results, that sometimes
differ by up to five sub-types. We attribute the poor agreement to the significant 
background contamination of the archival photometry -- both for old
photographic surveys, and the mid-infrared WISE and Neo-WISE surveys. 
We adopted as our final spectral types those obtained by the templates 
comparison, as the most direct and robust and  adopt uncertainties of 
their estimation equal one sub-type.  

\subsubsection{Spectral typing by comparison with spectral templates}

Nine of the twelve spectra showed the characteristics
molecular absorption bands of K and M type stars (TiO, CaH), and they
were compared to the primary K7V-M5V standards from Kirkpatrick \etal (1991, 1999)
available from the Dwarf Archives\footnote{http://www.dwarfarchives.org}.

The template spectra were smoothed to the resolution of our
data, normalised at 7500 \AA, and a $\chi^2$ minimisation over 
$\lambda$=5000--8000\,\AA\ was used to find the best match. The results are shown
in Fig.\,1. The types of some targets were adjusted by up to 0.5
sub-type after a visual inspection. The remaining three spectra
indicated hotter stars, and for those earlier than K7 type objects
we used the standards of Pickles (1998), following the same minimisation 
routine as before, but now over $\lambda$=3800--8000\,\AA. 
In case of VVV~J121051.57-642528.5, we cannot find a convenient fit  to  the  low resolution spectrum (see Fig.\,1). Target was overexpose and detector worked in nonlinear regime,
what prevented proper flux calibration.  Spectral type of this object was estimated by comparison of Na I doublet around 5890\,\AA~  in the high resolution spectrum with template spectrum. Both spectra were normalised by continuum. Fe I lines around Na I doublet fit perfectly what suggest similar rotation velocity.  
The derived spectral types are listed in Table\,2.

%===========================================================================
\begin{figure}
[htb]
  \includegraphics[width=125mm]{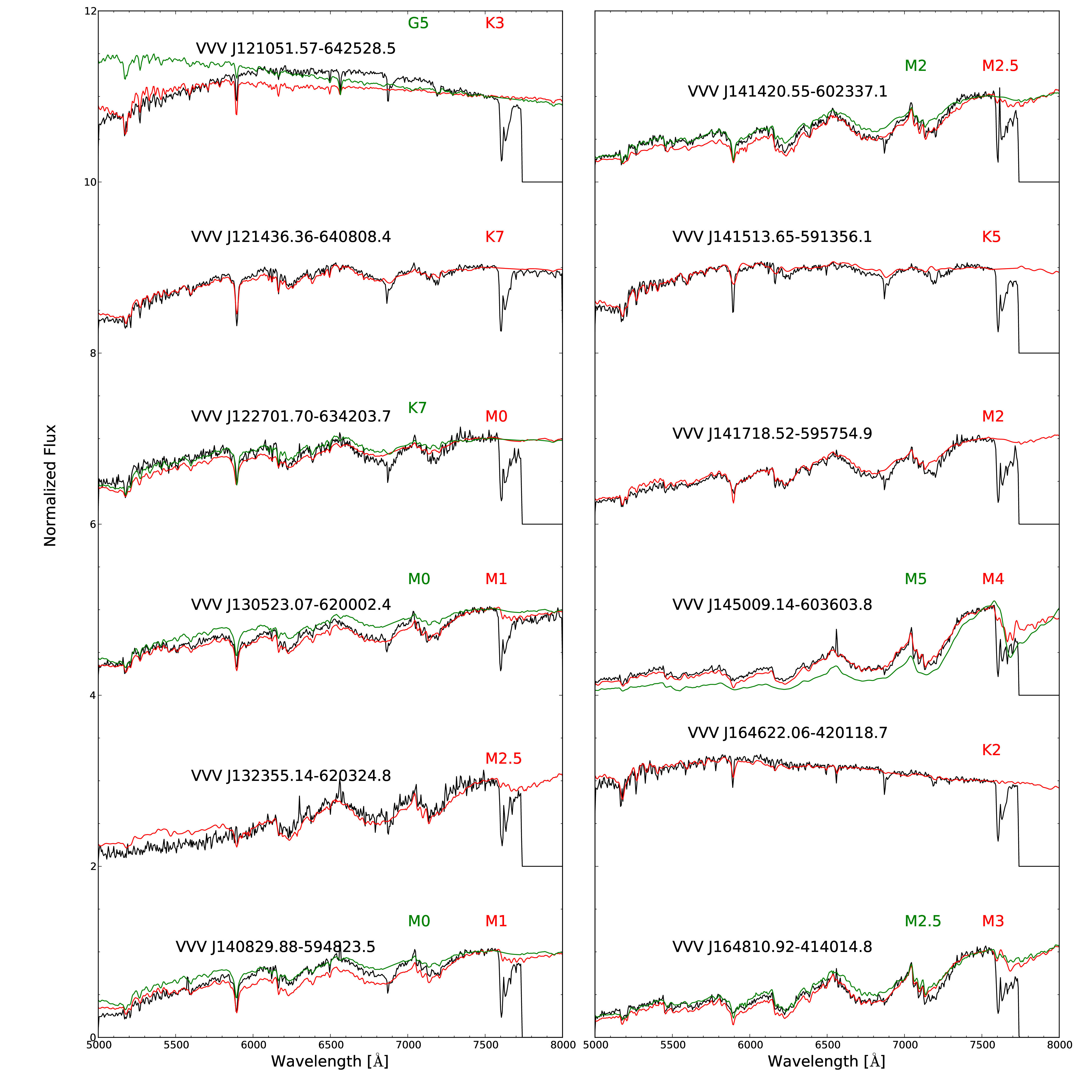}
\FigCap{
Observed spectra (black lines) overplotted with the best fitted template spectra (red lines) 
and alternative comparison template spectra (green lines). The template spectra were taken 
from Kirkpatrick \etal (1991) and  Pickles (1998) for objects with spectral types later, 
and earlier than K5, respectively. The absorption feature at $\sim$7600-7700\,\AA\
has telluric origin.}
\end{figure}
%===========================================================================

\subsubsection{Spectral typing using spectral indices}

Kirkpatrick \etal (1991) developed a system of
spectroscopic indices for mid-K to late-M stars, and calibrated them 
versus the spectral types. Other index systems have been developed 
afterwords, but we prefer this one because it has been applied to a 
large number of stars, providing broad comparison with literature 
data. We followed the definitions in their Table\,6, but measured 
only three diagnostic index ratios, because of our narrower spectral 
coverage. We compared the results with their Fig.\,6 to assign 
spectral types to the individual objects (Table\,1). 
Our errors are tentative, derived from the intrinsic spread of the 
``calibrators'' in Fig.\,6 of Kirkpatrick \etal (1991). Our formal 
Poisson errors are negligible, and we also verified that a small 
velocity offset of $\pm$20\,km\,s$^{-1}$ led to changes in the ratios 
of $\leq$0.5\%. 

For most stars the spectral types derived from different ratios agree 
well, with a exception VVV J140829.88-594823.5. To verify our 
classification we measured some K and M type spectra from the Dwarf 
Archives (Kirkpatrick \etal 1999), 
and successfully recovered the spectral classes of the test stars, 
typically with an uncertainty of 0.5-1 sub-types. However, ratio A 
appears to be the most reliable, because of the steeper calibration 
and wider dynamical range than the other ratios.

\MakeTable{lrlrlrl}{12.5cm}{Spectroscopic index ratios (as defined by Kirkpatrick \etal 1999), 
and derived spectral types.}
{
\hline\hline
Name~&~Ratio\,A~&~Sp.\,T.~&~Ratio\,B~&~Sp.\,T.~&~Ratio\,C~&~Sp.\,T.~\\     
(1)      & (2)~~~   & ~~(3)   & (4)~~~   & ~~(5)   & (6)~~~   & ~~(7)   \\
\hline
VVV J121051.57$-$642528.5 & 0.985~ & $<$K8         & 1.034~ & $\leqslant$K9 &        & \\       
VVV J121436.36$-$640808.4 & 1.081~ & M0.5$\pm$1.0  & 1.031~ & $\leqslant$M2 & 1.133~ & $\leqslant$M3 \\   
VVV J122701.70$-$634203.7 & 1.109~ & M0.5$\pm$1.0  & 0.957~ & $<$K8         &        & \\
VVV J130523.07$-$620002.4 & 1.103~ & M0.5$\pm$1.0  & 1.042~ & $\leqslant$M2 & 1.163~ & $\leqslant$M4 \\  
VVV J132355.14$-$620324.9 & 1.125~ & M1.0$\pm$1.0  & 0.924~ & $<$K8         & 1.101~ & $\leqslant$M2 \\ 
VVV J140829.88$-$594823.5 & 1.124~ & M1.0$\pm$1.0  & 1.075~ & M4.0$\pm$2.0  & 1.095~ & $\leqslant$M2 \\  
VVV J141420.55$-$602337.1 & 1.203~ & M2.5$\pm$0.5  & 1.050~ & $\leqslant$M3 &        & \\
VVV J141513.65$-$591356.1 & 1.034~ & $\leqslant$K9 & 1.044~ & $\leqslant$M3 &        & \\
VVV J141718.52$-$595755.0 & 1.120~ & M1.0$\pm$1.0  & 1.032~ & $\leqslant$M2 &        & \\
VVV J145009.14$-$603603.9 & 1.345~ & M4.5$\pm$1.0  & 1.081~ & M4.5$\pm$2.0  &        & \\
VVV J164622.06$-$420118.8 & 1.002~ & $<$K8         & 1.005~ & $\leqslant$K9 & 1.091~ & $\leqslant$M2 \\ 
VVV J164810.92$-$414014.9 & 1.220~ & M2.5$\pm$0.5  & 1.064~ & M2.5$\pm$2.0  &        & \\
\hline
}

\subsubsection{SED fitting} 

To produce the spectral energy distribution (SED) and compare to stellar models we used the
Virtual Observatory SED analyser (VOSA; Bayo \etal 2008), fitting the BT-settl 2012 models 
(Allard \etal 2012) and Kurucz ODFNEW /NOVER models (Castelli \etal 1997) to the data,
first with a Bayesian approach, and then constraining the model
parameters with a $\chi^2$ minimisation over three parameter space:
effective temperature $T_{\rm eff}$, surface gravity $\log g$, and metallicity
[Fe/H]. Increasing the number of photometric measurements and
widening the wavelength coverage makes the fit more robust and
reliable. Usually, the $T_{\rm eff}$ is the most stringently constrained
parameter, with uncertainties of order of $\sim$200\,K; The other 
parameters surface gravity and iron abundance, are usually, 
log\,g$\ge$4.5, and [Fe/H]$\ge$$-$1, with a typical scatter of 0.5\,dex for 
both. The photometry used in the SED fitting is described in Sec.\,2.4
and it is available in online materials.

%
%________________________________________________________________

\MakeTable{llcrrrrc}{12.5cm}{Spectral classifications (based on template comparison) and effective temperatures defined based on photometric spectral energy distribution, 
$J$-band magnitudes, proper motions (estimated based on 2MASS and the last available 
epoch of the VVV survey), spectro-photometric distances, and tangential velocities 
for the observed targets.}
{\hline\hline
Name   & Sp.\,T.   & SED $T_{\rm eff}$ & $J_{\rm 2MASS}$ & $\mu_{\alpha} \cos\delta$ & $\mu_{\delta}$ & Dist. & $V_{\rm tan}$  \\
       &           & [K]               & [mag]           &  [mas\,yr$^{-1}$]  & [mas\,yr$^{-1}$] &  [pc] & [km s$^{-1}$] \\
\hline
VVV J121051.57-642528.5 & K2   & 5300 &  8.211 & -118$\pm$10 &  -47$\pm$10 &   55 &  33$\pm$7  \\   
VVV J121436.36-640808.4 & K7   & 3900 &  9.424 & -181$\pm$9 &  -46$\pm$9 &   55 &  49$\pm$10  \\  
VVV J122701.70-634203.7 & K7   & 3700 & 12.742 &  178$\pm$10 &  -35$\pm$10 &  264 & 227$\pm$45  \\  
VVV J130523.07-620002.4 & M1   & 3700 & 11.851 & -169$\pm$8 &  -49$\pm$8 &  116 &  97$\pm$19  \\   
VVV J132355.14-620324.9 & M2.5 & 3600 &  9.865 &  227$\pm$9 & -200$\pm$9 &   35 &  50$\pm$10  \\
VVV J140829.88-594823.5 & M0   & 3600 & 10.340 & -162$\pm$11 &  -32$\pm$11 &   68 &  53$\pm$11  \\
VVV J141420.55-602337.1 & M2.5 & 3700 & 11.739 & -185$\pm$10 &  -65$\pm$10 &   84 &  78$\pm$16  \\
VVV J141513.65-591356.1 & K5   & 4000 & 10.974 & -120$\pm$8 & -143$\pm$8 &  144 & 127$\pm$25  \\
VVV J141718.52-595755.0 & M2   & 4000 & 11.113 & -133$\pm$8 &  -95$\pm$8 &   69 &  53$\pm$11  \\
VVV J145009.14-603603.9 & M4   & 3000 & 10.533 & -128$\pm$8 & -130$\pm$8 &   27 &  23$\pm$5  \\
VVV J164622.06-420118.8 & K2   & 4700 &  9.963 &  -46$\pm$9 &  -60$\pm$9 &  117 &  41$\pm$8  \\
VVV J164810.92-414014.9 & M3   & 3200 & 10.807 & -138$\pm$8 & -263$\pm$8 &   48 &  68$\pm$14  \\
\hline
}

\subsection{Distance estimation}

We were not able to measure parallaxes from the multi-epoch K$_{\rm S}$-band VVV data as achieved by
Beam\'in \etal (2013, 2015) and Smith \etal (2015). This is due to our targets being saturated on the VVV $K_{\rm s}$ images. Instead, we estimated their 
spectro-photometric distances from the spectral types we obtain (see Sec.\,3.1), and from the 2MASS $J$ and $K_{\rm S}$ photometry using the absolute magnitudes 
$M_{J}$, $M_{K_{\rm S}}$ from
Pecaut \& Mamajek (2013)\footnote{http://www.pas.rochester.edu/$\sim$emamajek/EEM\_dwarf\_UBVIJHK\_ colors\_Teff.txt}.
For each object we calculated the distances for the two 2MASS photometric bands separately.
Our final estimate was the average of these two measurements.
The mean difference is 1.6\%$\pm$0.8\%, with a maximum 
value of 3.1\% for object VVV\,J141513.65-591356.1.
The affect of the sub-type classification uncertainty on the distance estimation is of the order of 15\%, on average. The photometric uncertainties introduce 
uncertainties within 1--3\% in the magnitude range we examine. Finally, our estimated spectro-photometric distances should have an associated error of $\sim 20$\%. Derived distances are listed in Table\,2. 

\subsection{Co-moving wide binaries}

Six of our objects have co-moving companions.
We estimated the $T_{\rm eff}$ of the companions with SED fitting
(see Sec.\,3.1..3; the photometry is available in online materials
We also derived spectral types of companions using magnitude 
difference with the primary in $J$ band and the absolute 
magnitudes for a given spectral type from Pecaut \& Mamajek (2013).
The basic information about the binary systems are 
 summarised in Table\,3.

\MakeTable{lcccccccc}{12.5cm}{Binary systems investigated in this paper.
The columns are: spectral types from this work,
$T_{\rm eff}$ defined based on photometric spectral energy distribution, 
2MASS $J$-band magnitudes, angular separations in arcsec,
photometric distances in pc,
and proper motions in celestial coordinates.}
{\hline\hline
Name    &~Sp.\,T.    & SED $T_{\rm eff}$ & $~J_{\rm 2MASS}$ &~$\rho$ &~Dist. & ~$\rho_{\rm p}$ &~$\mu_{\alpha} \cos\delta$ &~$\mu_{\delta}$ \\     
        &            & [K]           & [mag]            & ['']  & [pc] & [a.u.]  & [mas\,yr$^{-1}$]                     & [mas\,yr$^{-1}$]    \\
\hline\hline
VVV J121051.57-642528.5 & K2 & 5300 & 8.211 & 45.8 & 55 & 2521 & -118$\pm$10 & -47$\pm$10 \\
VVV J121050.12-642443.5 & M5$^{\rm a}$ & 3100 & 13.064 & - & -  & - & -108$\pm$10 & -49$\pm$10  \\  
\multicolumn{8}{c}{}\\
VVV J121436.36-640808.4 & K7  & 3900 & 9.424  & 21.3 & 55 & 1170 & -181$\pm$9 & -46$\pm$9  \\ 
VVV J121433.35-640801.0 & M4$^{\rm a}$ & 3100 & 12.234 &  -  & - & - & -176$\pm$9 & -52$\pm$9  \\
\multicolumn{8}{c}{}\\
VVV J140829.88-594823.5 & M0 & 3600 & 10.340 & 16.7  & 68 & 1135 & -162$\pm$11 & -32$\pm$11  \\ 
VVV J140831.63-594834.3 & K8$^{\rm a}$  & 3600 & 10.024 & -  &   -   & - & -155$\pm$11 & -31$\pm$11  \\ 
\multicolumn{8}{c}{}\\
VVV J141420.55-602337.1 & M2.5 & 3700 & 11.739 & 11.7  & 84 & 983 & -185$\pm$10 & -65$\pm$10  \\
VVV J141421.23-602326.1 &  -   &  -   & 16.752 &   -  &  -  & - & -171$\pm$10 & -78$\pm$10  \\
\multicolumn{8}{c}{}\\
VVV J141718.52-595755.0 & M2 & 4000 & 11.113 & 2.8  & 69  & 193 & -133$\pm$8 & -95$\pm$8  \\
VVV J141718.30-595756.0 & M3$^{\rm a}$ &  -   & 11.845$^{\rm b}$ &  - &  -  & - & -143$\pm$8 & -105$\pm$8  \\
\multicolumn{8}{c}{}\\
VVV J164622.06-420118.8 & K2 & 4700  & 9.963 & 5.0  & 117 & 585  & -46$\pm$9 & -60$\pm$9  \\
VVV J164621.64-420119.0 & M0$^{\rm a}$  & 4300 & 11.520 &  -  & - & -   & -40$\pm$9 &  -61$\pm$9  \\
\hline
\multicolumn{9}{p{12.5cm}}{$^{\rm a}$ Sp. types of companions estimated base on magnitude differences and absolute magnitudes given by
Pecaut \& Mamajek (2013). $^{\rm b}$ System not resolved in achival images, magnitudes were taken from VVV catalogs and then transform 
to the 2MASS system.}
}
 
\subsection{A possible massive white dwarf near the ZZ~Ceti instability strip}

Finch \etal (2010) reported that the object 
UPM 1414$-$6023A  (=VVV~J141420.55$-$602337.1)
has a common PM companion  
UPM 1414$-$6023B (=VVV~J141421.23$-$602326.1).

This object is too faint and it is not detected by
2MASS, but it is relatively bright on the optical SuperCOSMOS plate,
suggesting that may be a white dwarf. We do not have a spectrum of this star.
We compiled a set of colours, including the VVV
$ZYJHK_{\rm s}$ magnitudes (transforming the $JHK_{\rm S}$ to the 2MASS system following Soto \etal 2013),
and the DENIS $I$-band, and compare them with the synthetic colour of pure hydrogen white dwarfs 
form Pierre Bergeron webpage\footnote{http://www.astro.umontreal.ca/$\sim$bergeron/CoolingModels}
(Holberg \& Bergeron 2006; Kowalski \& Saumon 2006; Tremblay \etal 2011; Bergeron \etal 2011).

The positions of VVV~J141421.23$-$602326.1 on the colour-colour and colour-magnitude diagrams indicate
effective temperature T$_{\rm eff}$$\sim$12\,000\,K and a relatively high mass of $\sim$1$M_{\odot}$
(Fig.~2). These parameters place it in the the ZZ\,Ceti instability strip. 
Typically, the ZZ~Ceti pulsators have $\log g\sim$8.2 (see Fig. 33 in Gianninas \etal 2011), that corresponds
to mass $\sim$0.6 $M_{\odot}$ so, VVV~J141421.23$-$602326.1 is unusually massive for this class of variables. 
There are only around 30 known stars more massive than 0.8 $M_{\odot}$ within the approved ZZ\,Cetis, and only three of them are more massive than 1$M_{\odot}$ (Castanheira \& Kepler 2014). 
Each new member of this tiny family is valuable, because the larger number of massive pulsators will allows to probe the ensemble internal structure of the high-mass end of 
the ZZ\,Ceti instability strip.  

Further spectroscopic and photometric follow-up is needed to confirm nature and to determine the age of this object.
High time resolution photometry could detect its variability and if it is indeed a ZZ~Ceti star, providing us with additional constraints about its structure and composition. 
 
%===========================================================================
\begin{figure}[htb]
  \includegraphics[width=125mm]{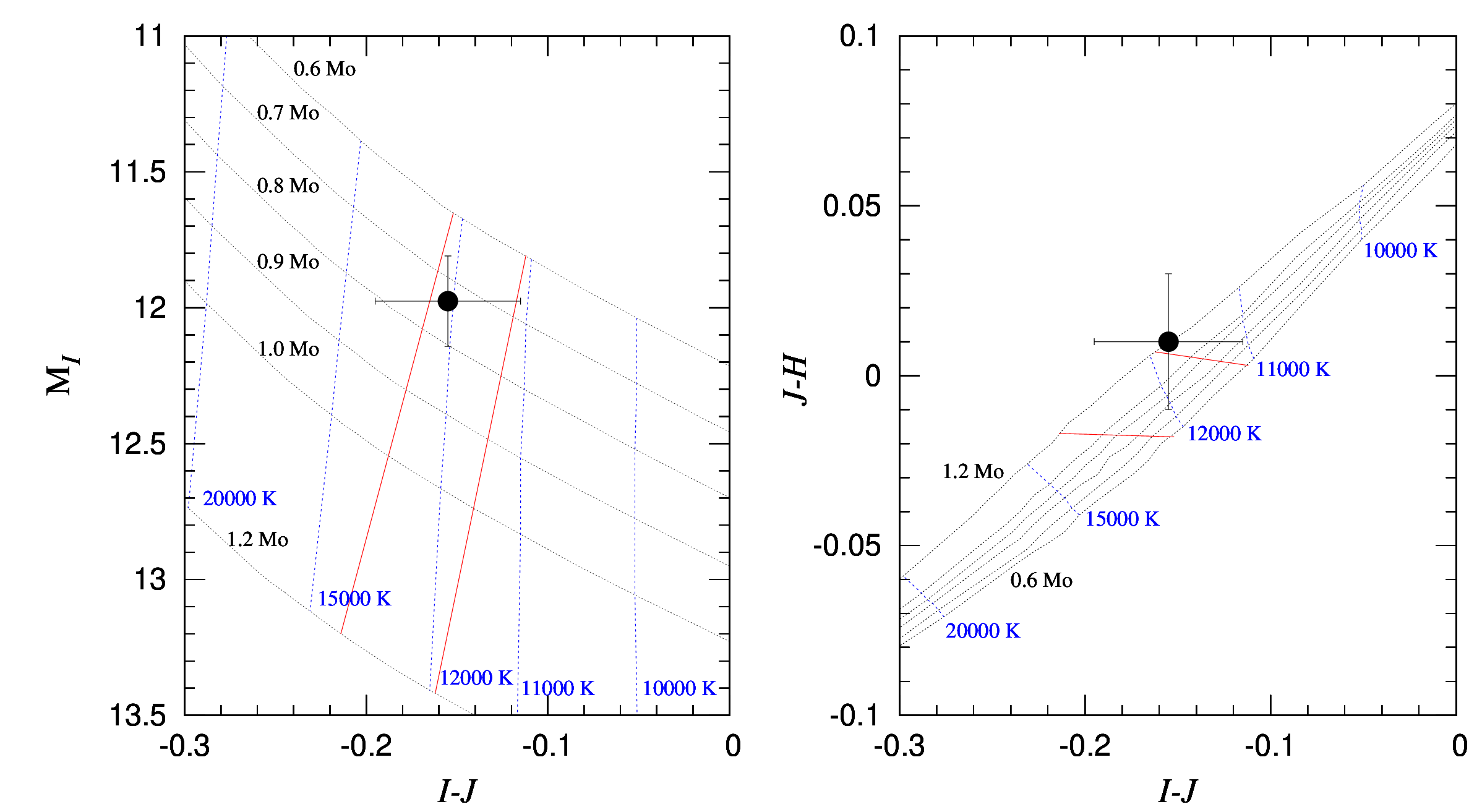}
\FigCap{
  Position of our new massive ZZ~Ceti type pulsator candidate, VVV~J141421.23-602326.1, 
  in the colour-magnitude diagram (left panel) and colour--colour diagram (right panel). 
  The black dot represents the measurements of VVV~J141421.23-602326.1 along with the 
  3-sigma errorbars. The black dashed lines represent the sythetic colours for a given 
  mass for a pure hydrogen atmoshpere model. The blue dashed lines show the model isotherms. The red solid 
  lines present borders of ZZ~Ceti instability strip taken from Gianninas \etal (2011).}
\end{figure}
%===========================================================================

\subsection{Kinematics}

We measured radial velocities for three objects: VVV~J121436.36-640808.4,  VVV~J132355.14-620324.9 and VVV~J164810.92-414014.9, 
using FEROS spectrograph at the MPG/ESP 2.2m telescope (Sec.\,2.2).
The $UWV$ galactic space velocities were computed with GAL\_UVW routine from the {\sc IDL} Astronomy User's Library.
The results are listed in Table\,4.  
We used the BANYAN II web tool\footnote{http://www.astro.umontreal.ca/$\sim$gagne/banyanII.php}  
(Gagn\'e \etal 2014; Malo \etal 2013) to estimate the probably that these three objects
may be members of any known young moving group. 
Typically, the derived spatial velocities are consistent with old field population:
100, 99.61, and 36.47\,\%, for VVV~J132355.14-620324.9, VVV~J164810.92-414014.9, and 
VVV~J121051.57-642528.5, respectively. 
There is a marginal possibility of 0.39\,\% that VVV~J164810.92-414014.9 may belong to 
the young field. This possibility is considerable for VVV~J121051.57-642528.5 - 33.60\,\%.
Finally, there is 29.81\,\% probability that the last object may belong to the Argus moving
group and marginal chance of 0.11\,\% that it may belong to $\beta$ Pictoris moving group.
Although, high resolution spectrum of this target does not show Li I doublet at $\sim$6708\,\AA, 
what suggests that its age is higher than 150 Myr and excludes membership to any know young moving group. 

\MakeTable{lrrrr}{12.5cm}{Radial velocities RV and $UVW$ galactic space velocities. 
  All velocities are expressed in km s$^{-1}$.}
{\hline\hline
VVV Name   & RV & $U$ & $V$ & $W$  \\
\hline
J121051.57-642528.5 &   3.27$\pm$0.01 & -23.7$\pm$3.0 & -16.0$\pm$1.6 & -17.0$\pm$2.0 \\
J132355.14-620324.9 & 123.20$\pm$0.03 &  100.4$\pm$5.4 & -79.0$\pm$3.9 & -36.4$\pm$7.5 \\
J164810.92-414014.9 & -35.08$\pm$0.09 & -52.4$\pm$3.6 & -52.9$\pm$11.9 & -15.8$\pm$2.7 \\
\hline}

\section{Summary and conclusions}

We obtained spectroscopic follow-up observations of twelve new high PM objects found by the VVV survey during the initial testing of our searching method, and we also looked for possible new wide binary companions. We derived their optical spectral types and photometric distances.

All of the analysed objects are K and M dwarfs located at 27--264\,pc from the Sun and are bright enough for further follow-up and search of planets using state of the art and upcoming NIR instruments. Also, all objects are in the densest regions of the Milky Way, surrounded by a pletora of bright stars, very suitable for AO imaging. That makes our targets ideal for searches of close neighbours. From the other side, the surrounded stars are ideal comparison stars for precise relative photometry, variability and transit studies.

VVV~J141421.23-602326.1, a co-moving companion of VVV~J141420.55-602337.1, is a candidate for 
being a rare massive ZZ~Ceti type pulsator. Further spectroscopic and photometric follow-up is needed 
to better constrain nature and age of this object.

\Acknow{We gratefully acknowledge use of data from the ESO Public Survey programme ID 179.B-2002
taken with the VISTA telescope, and data products from the Cambridge Astronomical Survey
Unit. 
This publication makes use of data products from the Two Micron All Sky Survey, which is a joint project of the University 
of Massachusetts and the Infrared Processing and Analysis Center/California Institute of Technology, funded by 
NASA and NSF.This research has benefitted from the M, L, T, and Y dwarf compendium housed at DwarfArchives.org. 
Support for MG, RK, JCB, DM, and JB is provided by the Ministry of Economy,
Development, and Tourisms Millennium Science Initiative through grant IC120009, awarded to The
Millennium Institute of Astrophysics, MAS.
MG acknowledges support from Joined Committee ESO and Government of Chile 2014.
RK, DM and JB are supported by FONDECYT grants No. 1130140, 1130196 and 1120601, respectively.
Both PV and AYK acknowledge support from the National Research Foundation of South Africa.
JB, RK, MG and VV are supported by CONICYT REDES140042. JCB acknowledge support from CONICYT FONDO GEMINI - Programa de Astronom{\'i}a del DRI, Folio 32130012.}

\end{document}